\RequirePackage{lineno}
\documentclass[twocolumn,letterpaper,aps,prc,superscriptaddress,showpacs,floatfix]{revtex4-1}

\usepackage{graphicx}	
\usepackage{subfigure}

\graphicspath{%
  {./},%
  {./figures/}%
}

\usepackage{xspace}	
\usepackage{color}

\newcommand{\pt}{\mbox{$p_T$}\xspace}

\newcommand{\sqsn}{\mbox{$\sqrt{s_{_{NN}}}$}\xspace}
\newcommand{\dau}{\mbox{$d$+Au}\xspace}
\newcommand{\pau}{\mbox{$p$+Au}\xspace}
\newcommand{\pal}{\mbox{$p$+Al}\xspace}
\newcommand{\hau}{\mbox{$^3\text{He}$+Au}\xspace}
\newcommand{\pp}{\mbox{$p$+$p$}\xspace}
\newcommand{\ppb}{\mbox{$p$+Pb}\xspace}

\begin{document}
\title{Azimuthal Anisotropy Relative to the Participant Plane from \textsc{AMPT} in Central $p+$Au, $d+$Au, and $^3$He+Au Collisions at \sqsn = 200 GeV }

%

\author{J.D. Orjuela Koop, A. Adare, D. McGlinchey, J.L. Nagle}
\affiliation{University of Colorado, Boulder}

\date{\today}

\begin{abstract}
Recent data from \pp and \ppb collisions at the Large Hadron Collider (LHC), and \dau and \hau collisions at the Relativistic Heavy Ion 
Collider (RHIC) reveal patterns that---when observed in the collision of heavy nuclei---are commonly interpreted as indicators of a locally equilibrated system in collective motion.  The comparison of these data sets, including the forthcoming 
results from \pau and \pal  collisions at RHIC, will help to elucidate the geometric dependence of such patterns. 
It has recently been shown that
A-Multi-Phase-Transport-Model (\textsc{AMPT}) can describe some of these features in LHC data with a parton-parton scattering
cross section comparable to that required to describe $A+A$ data. In this paper, we extend these studies by
incorporating a full wave function description of the $^3$He nucleus to calculate elliptical and triangular anisotropy moments $v_2$ and $v_3$ for \pau, \dau and \hau collisions at the RHIC top energy of 200 GeV. We find reasonable agreement with 
the measured $v_2$ in \dau and \hau and $v_3$ in \hau for transverse momentum (\pt) $\lesssim$ 1 GeV/c, but underestimate these measurements for higher values of \pt. We predict a pattern of coefficients ($v_{2}$, $v_{3}$) for \pau, dominated by differences in the number of induced local hot spots (i.e. one, two, or three) arising from intrinsic geometry. Additionally, we examine how  this substantial azimuthal anisotropy accrues during each individual evolutionary phase of the collision in the \textsc{AMPT} model.
The possibility of a simultaneous description of RHIC- and LHC-energy data, the suite of different geometries, 
and high multiplicity \pp data is an exciting possibility for understanding the underlying physics in these systems.   

\end{abstract}


\maketitle

\section{Introduction}
Recent analyses of data from \pp and \ppb collisions at the LHC, as well as from \dau and \hau at RHIC have revealed the existence of azimuthal particle correlations reminiscent of those observed in $A+A$ collisions~\cite{adare_measurement_2014,Adare:2015ita,atlas_observation_2012,alice_long_2013,cms_observation_2012,khachatryan_observation_2010}. In the latter, this signal has been interpreted as evidence for the liquid-like nature of the Quark-Gluon Plasma (QGP)---a locally equilibrated, strongly coupled medium undergoing hydrodynamic expansion with viscosity near the quantum lower bound. However, these results were largely unexpected in $p(d)+$A, long considered control systems 
to understand initial-state effects in heavier systems. 

The success of nearly inviscid hydrodynamics in describing $A+A$ bulk observables makes it natural to ask if droplets of QGP are being formed in small systems, and if the created medium is sufficiently long lived to equilibrate locally and translate initial spatial anisotropies into final-state particle momentum correlations. However, this is not the only possibility, with potentially different physics being able to account for these measurements~\cite{bzdak_elliptic_2014,ma_long-range_2014,dusling_azimuthal_2012}. Further insight into this matter will come from the confrontation of different model calculations with the full data sets available.

It has been found that azimuthal two-particle correlations from high multiplicity \pp, \ppb, \dau, and \hau events exhibit an
enhancement around $\Delta \phi \approx 0$ (i.e. near-side), even when the particles have a large separation in 
pseudo-rapidity ($\Delta \eta > 2$), where
jet contributions are expected to be minimal.  There is an additional enhancement from \pp and peripheral \ppb(\dau)
to central \ppb(\dau) around $\Delta \phi \approx \pi$ (i.e. away-side) that has been interpreted as the full 
azimuthal continuation of elliptical and triangular flow coefficients, $v_2$ and $v_3$.   Alternative readings 
of the away-side pattern include modification of the dijet correlations for the most central \dau 
collisions~\cite{Adamczyk:2014fcx}.

Predictions for LHC-energy \ppb anisotropies using nearly inviscid hydrodynamics~\cite{bozek_collective_2012} provide a reasonable description
of the flow coefficients measured at the LHC.   However, as expected, an exact quantitative description depends on 
on the shear viscosity to entropy density ratio ($\eta/s$) and the details
of how initial geometry is modeled, for which there are quite different possibilities in $p+A$ collisions---see
for example Ref.~\cite{Schlichting:2014ip,Bzdak:2013zma}.  
There are also competing calculations where final-state QGP or flow effects
are deemed negligible, and it is initial-state glasma diagrams that give rise to the correlations~\cite{dusling_azimuthal_2012}.   In
\dau and \hau collisions, the initial geometry is dominated by the spatial separation of the two nucleons in the deuteron, reducing
differences between models of geometry.  For this case, nearly inviscid hydrodynamic calculations give a reasonable description of the 
experimentally extracted flow coefficients~\cite{nagle_exploiting_2013,Bzdak:2013zma,Bozek:2014xsa}.    

However, questions regarding the validity of the near-inviscid hydrodynamic calculations have been raised in terms of the expansion around steep energy density gradients in these small systems~\cite{Romatschke:2015dha,Romatschke:2015gxa,Niemi:2014wta,Denicol:2015bpa}. It is thus quite interesting that incoherent elastic parton-parton scattering---as implemented in A-Multi-Phase-Transport-Model (\textsc{AMPT})~\cite{lin_multiphase_2005}---can effectively reproduce the long-range azimuthal correlations~\cite{ma_long-range_2014} and $v_2$ coefficients~\cite{bzdak_elliptic_2014} observed in high multiplicity \pp and \ppb events at the LHC. Notice, however, that in the case of $v_2$ in \ppb, a good reproduction of the measured values is only achieved for \pt $\lesssim$ 2 GeV/c, above which the calculations underestimate the data. These \textsc{AMPT} results were obtained using a parton scattering cross section of $\sigma=1.5-3.0$ mb, and incorporating the so-called \textit{string melting} mechanism in the model (thus including a time stage dominated by parton-parton scattering). 

These results raise the question of whether a similar description can be achieved for different collision geometries
at the RHIC energy scale. In particular, we use \textsc{AMPT} to simulate \pau, \dau and \hau at $\sqsn = 200 \text{ GeV}$ since they have been proposed as an excellent testing ground to disentangle the properties of the medium created in small collision systems~\cite{nagle_exploiting_2013}. In this paper, we begin by describing the \textsc{AMPT} model and the methodology used to compute azimuthal anisotropies of final state hadrons with respect to the participant plane. We then compare our $v_2$ and $v_3$ results in \dau and \hau to available data, and present predictions for $v_2$ and $v_3$ in \pau. Finally, we discuss our results and provide some conclusions.

\section{Methods}

The \textsc{AMPT} event generator~\cite{lin_multiphase_2005} is a useful tool for the study of heavy-ion collision dynamics. The \textsc{AMPT} model
uses the HIJING model~\cite{wang_hijing_1994} just to generate initial conditions via Monte Carlo Glauber, Zhang's Parton  Cascade  (ZPC)  to  model  partonic  scattering, and  A-Relativistic-Transport (ART)  to  model late  stage  hadronic  scattering.   We  utilize  the  \textsc{AMPT} model with string melting turned on, such that a stage with parton-parton scattering is included and subsequent hadronization is described via a coalescence model. In this coalescence model, quark-antiquark pairs and sets of three (anti) quarks in close spatial proximity are grouped to form mesons and baryons, respectively.

In order to better understand how \textsc{AMPT} translates anisotropies in the initial geometry to anisotropies in final-state momentum, we modified the internal Monte Carlo Glauber to more closely resemble the standard approach described in~\cite{Miller:2007ri,Loizides:2014vua}, and used in~Ref.\cite{nagle_exploiting_2013}. The position of nucleons in each colliding nucleus is sampled from the appropriate wave function on an event-by-event basis, after which a nucleon-nucleon inelastic cross section of 42 mb is used to geometrically determine which nucleons were wounded in the collision. In the case of the deuteron, coordinates are sampled from the two-nucleon Hulth\'en wavefunction; in the case of $^3$He, coordinates are obtained with Green's function Monte Carlo calculations using the AV18+UIX model of three-body interactions~\citep{carlson_structure_1998}.  

We have run approximately 10 million central (i.e. impact parameter $b < 2$ fm) \textsc{AMPT} events with a parton-parton scattering cross section $\sigma = 1.5$ mb for each system at \sqsn = 200 GeV.   There are numerous publications utilizing the \textsc{AMPT} model to describe $A+A$ collisions at RHIC, quoting a full range of input cross section values ranging from $\sigma = 1.5$ mb up to $\sigma = 10$ mb~\cite{PhysRevC.86.054908,JPhysG30.2004.S263-S270,ChinesePhysC37.014104}. For this study, we have chosen the smallest value from this range to understand how a minimum parton scattering stage contributes to collective motion in these small systems. Table \ref{tab_partproduction} summarizes the mean number of nucleon participants for each system in \textsc{AMPT}, as well as the corresponding yield of partons at the end of the parton scattering stage and the yield of final-state hadrons at the end
of the hadronic scattering stage.   Note that all partons reported by \textsc{AMPT} at the end of the parton scattering stage are
quarks and anti-quarks, and no gluons, which are then input to the particular coalescence calculation for hadronization.

\begin{table}[tbh]
\caption{\label{tab_partproduction}  Particle production and eccentricity in central \textsc{AMPT} small-system collisions. For each collision system, we show the mean number of participant nucleons per event, the mean number of partons (i.e. quarks and antiquarks) at freeze out, the mean number of hadrons after the hadron cascade, and the mean elliptical and triangular initial state eccentricities.}
\begin{ruledtabular}
\begin{tabular}{cccccc}
  System & $\langle N_{part}\rangle$ &  $\langle N_{partons} \rangle$ & $\langle N_{hadrons}\rangle$ & $\langle\varepsilon_2 \rangle$ & $\langle \varepsilon_3 \rangle$ \\\hline
   \pau  &  10.45   &  182 &  131 & 0.24 & 0.16    \\
   \dau  &  18.3   &  336  &  233  & 0.57 & 0.17  \\
   \hau &  22.3   &  446  &  326  & 0.48 & 0.23  \\
    \end{tabular}
     \end{ruledtabular}
\end{table}

For each collision system, we compute the eccentricity $\varepsilon_n$, and participant plane angle $\Psi_n$ on an event-by-event basis using the initial-state coordinates $(r_i,\phi_i)$ of participant nucleons with the spatial distribution of a Gaussian of width $\sigma=0.4$ fm, as follows for $n=2,3$

\begin{equation}
\varepsilon_n = \frac{\sqrt{\langle r^2\cos(n\phi) \rangle^2 + \langle r^2\sin (n\phi) \rangle^2}}{\langle r^2\rangle},
\end{equation}

\begin{equation}
\Psi_n = \frac{\text{atan2}(\langle r^2 \sin (n\phi) \rangle,\langle r^2 \cos (n\phi) \rangle)}{n} +\frac{\pi}{n}.
\end{equation}
Average values of $\varepsilon_{2}$ and $\varepsilon_{3}$ for central \pau, \dau, and \hau are shown in Table~\ref{tab_partproduction}.

Having measured $\Psi_n$ from the initial-state geometry, we compute the second and third order azimuthal anisotropy moments $v_2$ and $v_3$ of final-state unidentified charged hadrons within $|\eta| < 2$, with respect to the participant planes, as follows
\begin{equation}
v_n = \langle \cos[n(\varphi-\Psi_n)] \rangle.
\end{equation} 

In addition to extracting anisotropy moments with respect to the participant plane, we are also interested in qualitatively examining the long-range azimuthal correlations of these hadrons. To that end, we follow an analysis procedure similar to that put forth by the ATLAS experiment for \ppb collisions in Ref.~\cite{atlas_observation_2012}.  We take all final-state charged particles and  consider all pairs separated by $2.0 < |\Delta \eta| < 3.0$ within a common \pt bin. We then form a long-range two-particle azimuthal correlation function for each \pt bin:

\begin{equation}
\label{C}
C(\Delta\phi,\pt) = \frac{1}{N_{\text{trig}}}\frac{dN(\pt)}{d\Delta\phi}.
\end{equation}  
 
\begin{figure}
\centering
\includegraphics[scale=0.55]{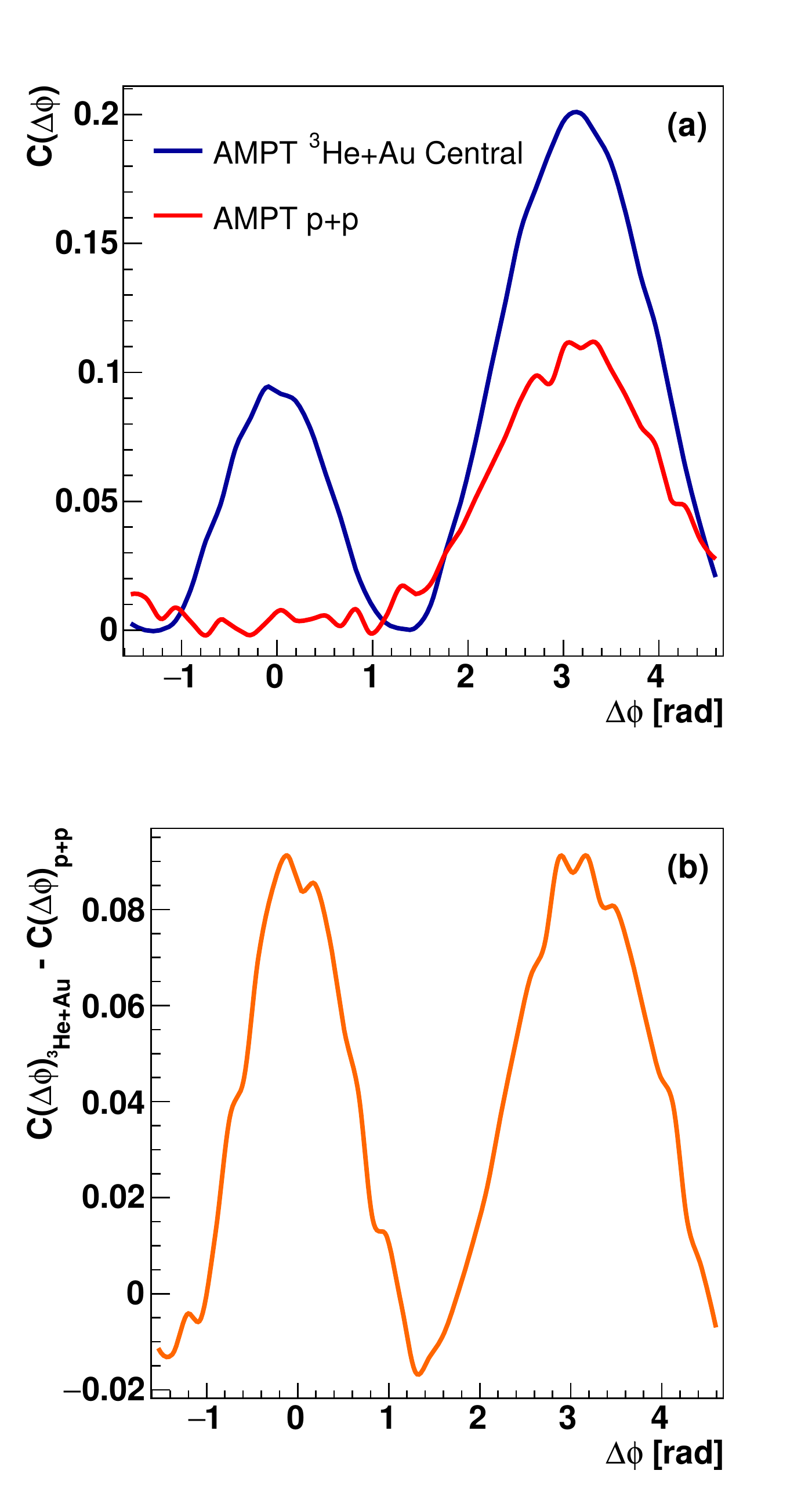}
\caption{(a) Two-particle correlation for charged particles within $0.90<p_T<1.04$ GeV/c for \pp and \hau 
events at \sqsn = 200 GeV. (b) Contributions to the correlation function arising from jet fragmentation are 
removed by subtracting away the per-trigger yield from \pp events. The resulting correlation function is shown in yellow.}
\label{fig_jet_contr}
\end{figure}

Shown in the top panel of Figure~\ref{fig_jet_contr} are the two-particle correlations from \pp and central \hau collisions for particles within $0.9<\pt<1.04$ GeV/c. As expected, a flat near-side (around $|\Delta\phi | \approx 0$) distribution is observed in \pp since two particles from the same jet fragmentation or resonance decay are very
unlikely to be separated by more than two units of pseudo-rapidity. There is also a significant \pp away-side enhancement
(around $|\Delta\phi | \approx \pi$) from jets back-to-back at leading order in azimuth and with a significant pseudo-rapidity
swing between them from an imbalance in the initial parton momentum fractions $x_{1}$ and $x_{2}$.   

In contrast, in the \hau distribution we observe a prominent near-side peak and a nearly two-fold enhancement of the away-side yield relative to \pp. Assuming the jet contribution to be the same in \pp and \hau, we subtract the two distributions, thus obtaining the result shown in the bottom panel of Figure~\ref{fig_jet_contr}. We choose to use \pp data for the jet subtraction, analogously to what has been done in experimental data with peripheral \ppb(\dau) events~\cite{atlas_observation_2012,alice_long_2013,PhysRevLett.111.212301}.
It is important to note that we do not use long-range two-particle correlations to extract anisotropy moments, but rather to provide a qualitative comparison with experimental results where similar correlations are presented. Hence, all results presented in the remainder of this article are computed with respect to the participant plane using initial state geometry, as previously described.


\section{Results}

\begin{figure}[ht]
\centering
\includegraphics[scale=0.45]{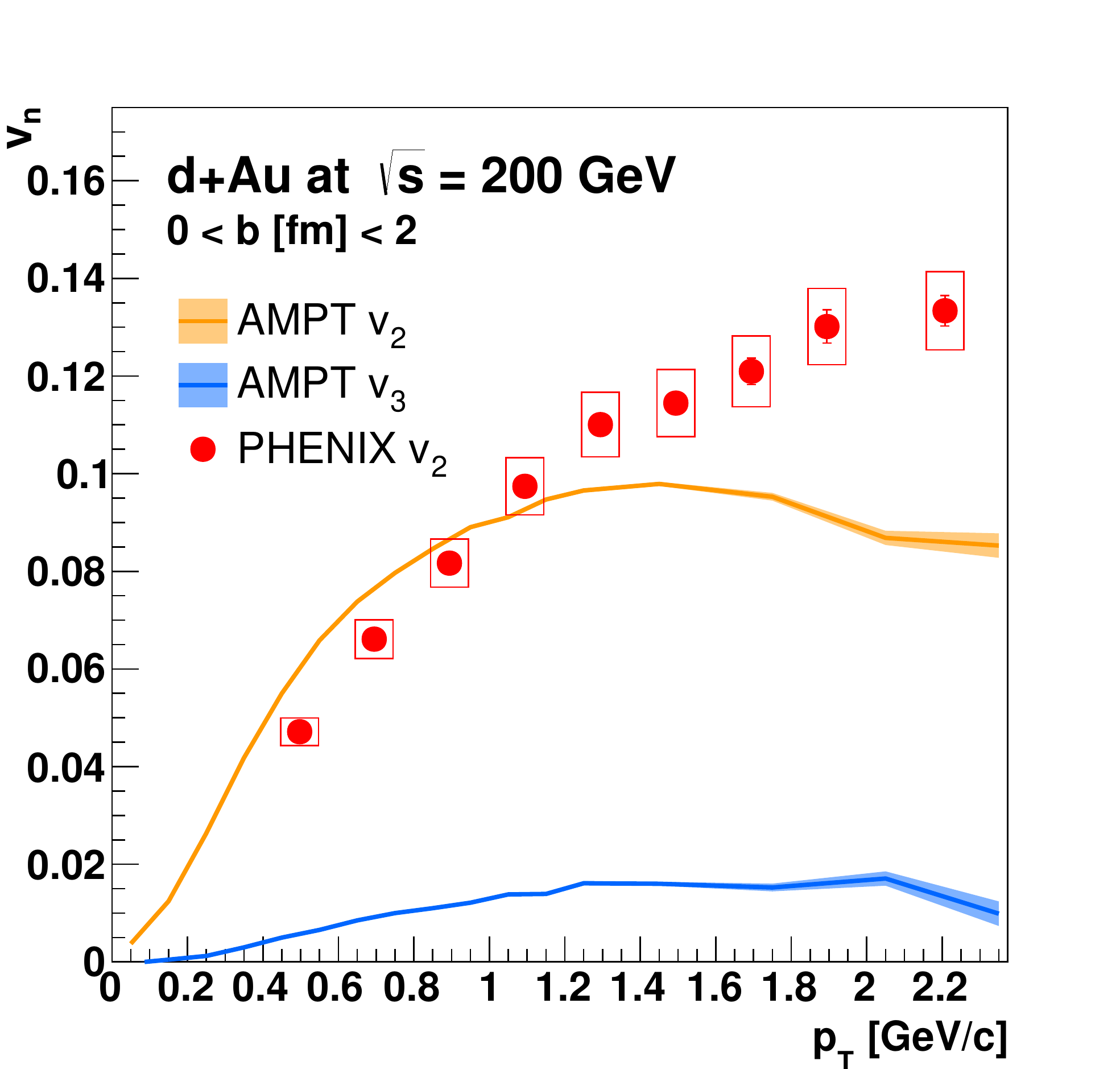}
\caption{\textsc{AMPT} calculation results for the azimuthal anisotropy moments $v_2$ and $v_3$ for central \dau at \sqsn = 200 GeV,
compared with experimental results from the PHENIX experiment.}
\label{fig_v2_dau}
\end{figure}

The resulting anisotropy moments $v_{2}$ and $v_{3}$ for \textsc{AMPT} \dau central collisions, computed with respect to the participant plane as described in Section II, as a function of \pt 
are shown in Figure~\ref{fig_v2_dau}.   We observe a substantial $v_{2}$ that rises with \pt to
around 10\% at \pt $\approx$ 1.0 GeV/c, after which it levels off and even exhibits a slight decrease.  
The $v_{3}$ coefficients exhibit a similar \pt dependent trend, though substantially smaller values.
Shown for comparison are elliptic flow measurements using the participant plane method 
for central (0-5\%) \dau events at $\sqsn = 200 \text{ GeV}$
from the PHENIX experiment~\citep{adare_measurement_2014}. There is a reasonable agreement between the \textsc{AMPT} calculation 
and the $v_2$ data below \pt $\approx$ 1.0 GeV/c, above which our calculation underestimates the experimental measurements.

\begin{figure}[ht]
\centering
\includegraphics[scale=0.45]{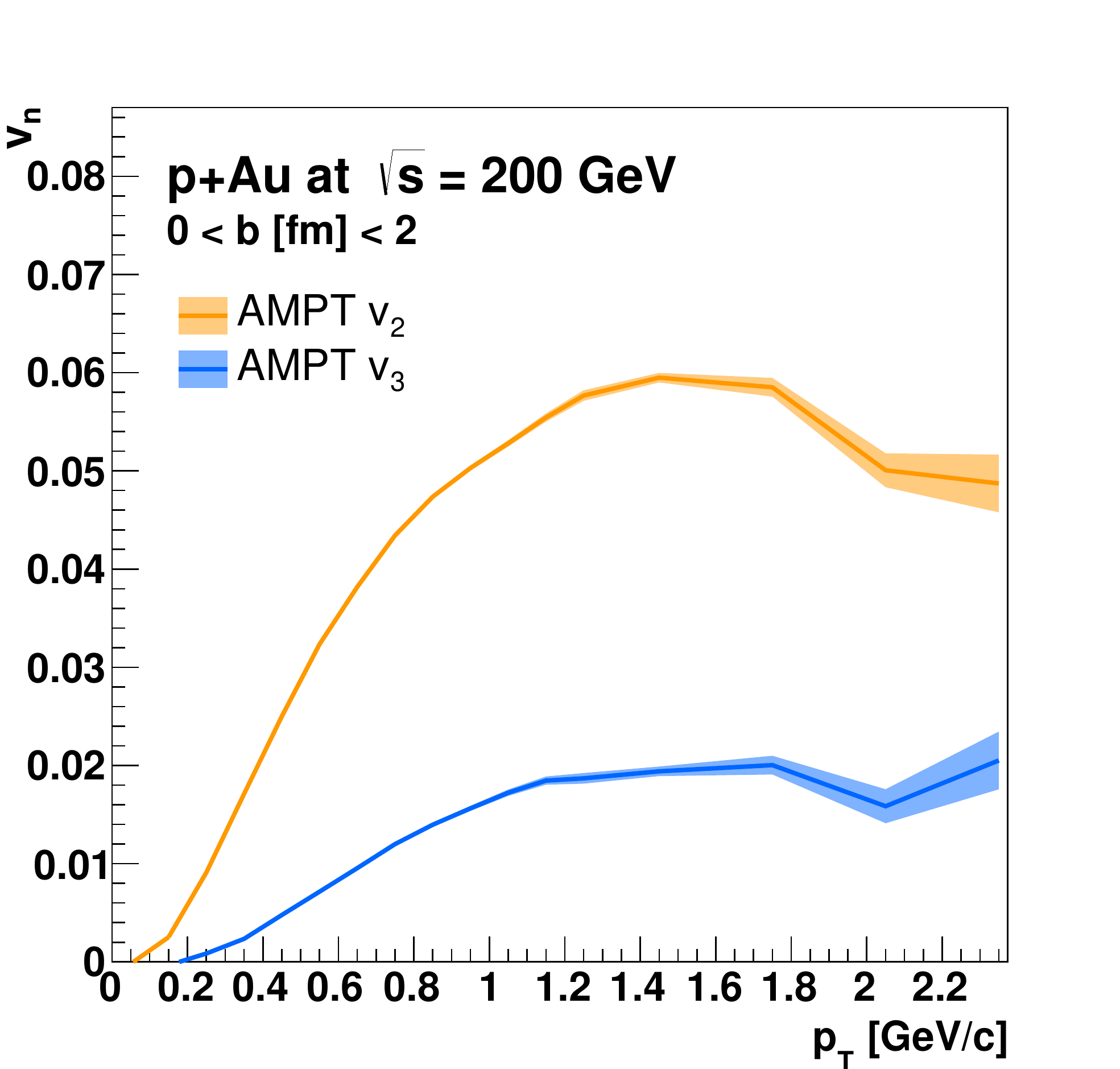}
\caption{\textsc{AMPT} calculation results for the azimuthal anisotropy moments $v_2$ and $v_3$ for central \pau at \sqsn = 200 GeV.}
\label{fig_vn_pau}
\end{figure}

\begin{figure}[ht]
\centering
\includegraphics[scale=0.45]{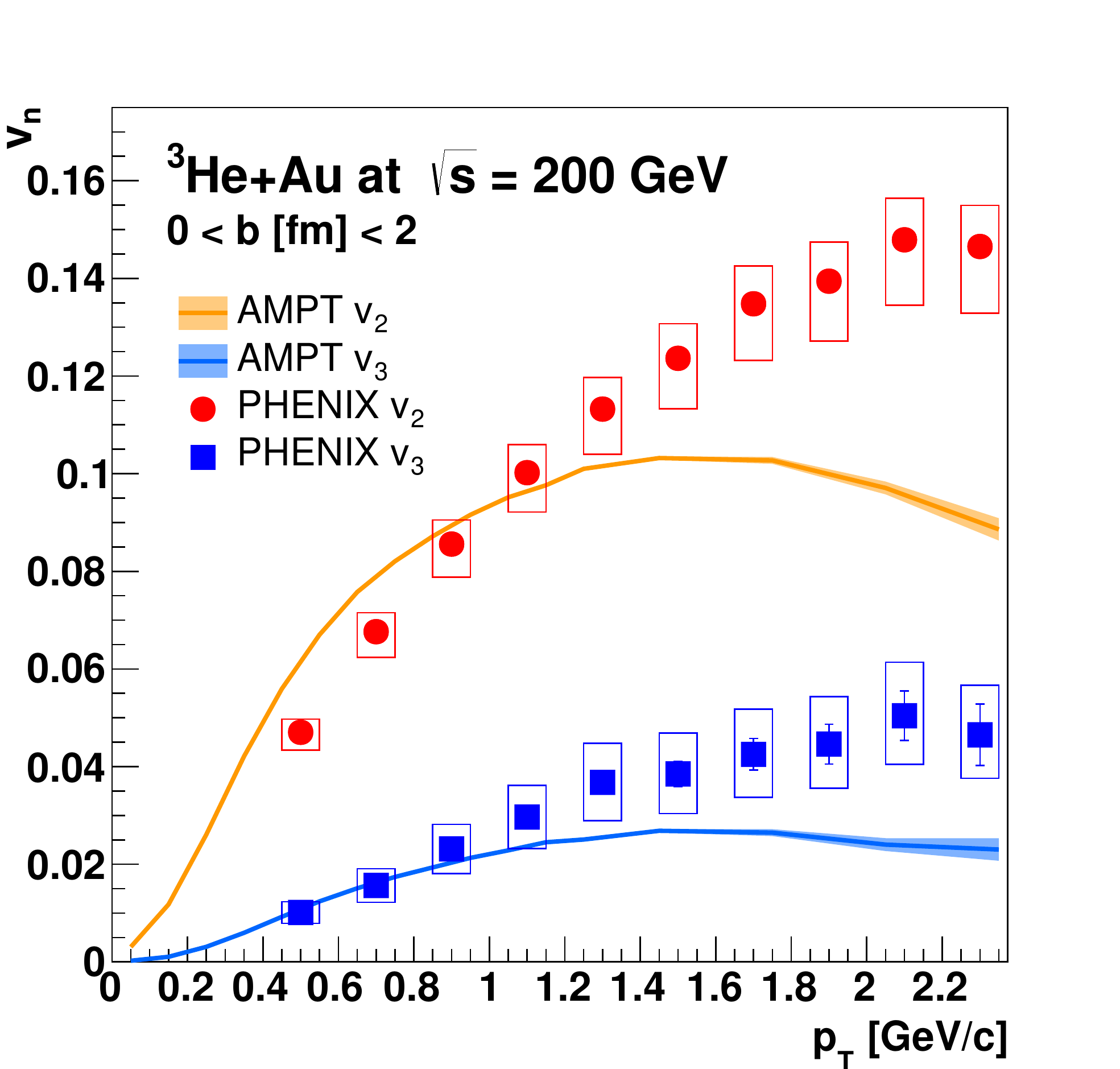}
\caption{\textsc{AMPT} calculation results for the azimuthal anisotropy moments $v_2$ and $v_3$ for central \hau at \sqsn = 200 GeV,
compared with experimental results from the PHENIX experiment.}
\label{fig_vn_he3au}
\end{figure}

Leaving all \textsc{AMPT} parameters fixed, we show in Figures~\ref{fig_vn_pau} and \ref{fig_vn_he3au} the
$v_2$ and $v_3$ anisotropy moments as a function of \pt for central \pau and \hau, respectively. 
The same general pattern of rising $v_{2}$ with \pt and smaller $v_{3}$ coefficients are observed for all systems.  
It is notable that there is an inflection point at $\pt \approx 1.5$ GeV/c, after which both the $v_{2}$  and $v_{3}$ exhibit a slight decreasing trend.

The RHIC experiments completed taking \hau collision data at \sqsn = 200 GeV in 2014.   
The PHENIX collaboration has presented $v_2$ and $v_3$ 
measurements in \hau at \sqsn = 200 GeV~\cite{Adare:2015ita}. 
These experimental results are reproduced in Figure \ref{fig_vn_he3au}, and are found to be in reasonable 
agreement with the \textsc{AMPT}-extracted coefficients up to \pt $\approx 1.0 \text{ GeV/c}$, beyond which the \textsc{AMPT} results fall 
significantly below the data, substantially more for $v_3$ than for $v_2$ if we consider the relative difference.


\section{Discussion}
Having obtained $v_2$ and $v_3$ for collision systems with different initial geometry and having compared them to experimental data, the question is then how to understand these results physically. In hydrodynamic models there is a straightforward physical picture of momentum anisotropy in terms of the velocity field of an expanding fluid. However, the situation is less clear in the case of transport models, such as \textsc{AMPT}. It has recently been proposed that the origin of the substantial $v_2$ calculated with \textsc{AMPT} lies predominantly in the anisotropic probability of partons to escape from the partonic scattering phase; that is, there is a preferential direction along which to freeze out~\cite{He:2015hfa}. In this section, we compare our elliptical and triangular anisotropy moments to understand how they relate to intrinsic initial geometry in \textsc{AMPT}.

\begin{figure}[ht]
\centering
\includegraphics[scale=0.5]{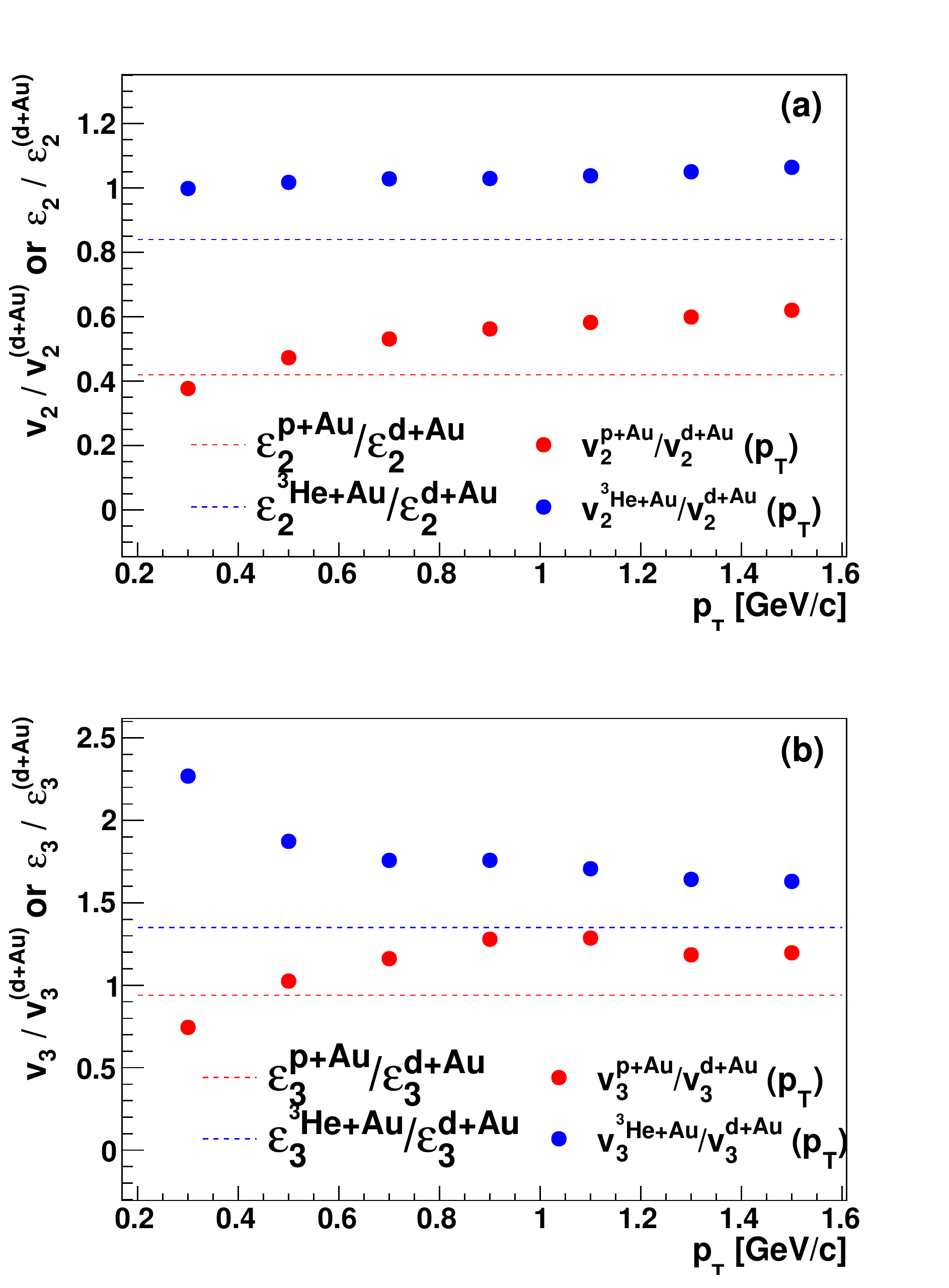}
\caption{(a) Ratio of elliptic and (b) triangular anisotropy moments as a function of \pt in \pau and \hau 
compared with a baseline in \dau.   Dashed lines are the ratios of $\varepsilon$ values from the initial geometry.}
\label{fig_vn_ratio}
\end{figure}

\begin{figure*}[ht]
\centering
\includegraphics[scale=0.8]{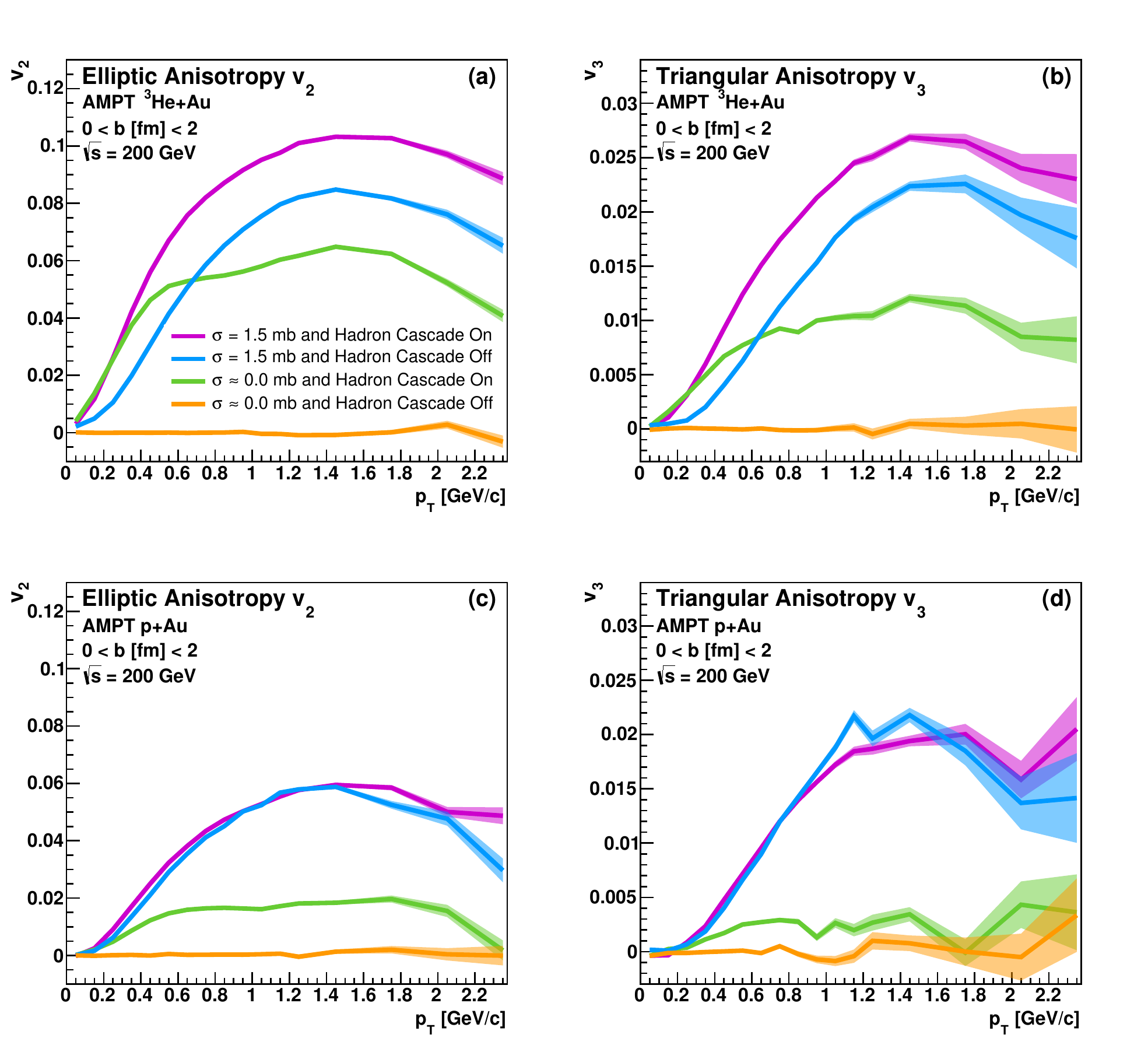}
\caption{Effects on $v_n$ of enabling and disabling the parton scattering phase and the hadronic cascade in \textsc{AMPT} for \hau---panels (a) and (b), and \pau---panels (c) and (d).}
\label{fig_cascade}
\end{figure*}

In hydrodynamic modeling, the final momentum anisotropies are directly related to the initial state eccentricities.
In order to explore whether this relationship holds in \textsc{AMPT}, we take the ratios of \textsc{AMPT} calculated $v_2$ and $v_3$ values
between systems.
Figure \ref{fig_vn_ratio} (top panel) shows the ratio of $v_2$ values as a function of \pt in 
\pau and \hau relative to \dau.   Figure \ref{fig_vn_ratio} (bottom panel) shows the ratio of $v_3$ values
 as a function of \pt in \pau and \hau relative to \dau.   Also shown are the ratios of initial eccentricities, 
$\varepsilon_{2}$ and $\varepsilon_{3}$ from Table~\ref{tab_partproduction}, as dashed lines.    The \textsc{AMPT}
$v_2$ and $v_3$ ratios between different systems are relatively \pt independent, with notable deviations at lower $\pt < 0.6$
GeV/c, and following the same ordering as the initial geometry ratios.    The values for the $\varepsilon$ ratios are all
lower than the \textsc{AMPT} $v_{n}$ ratios by approximately 15-35\%.   It is notable that the initial spatial eccentricities
are calculated from the nucleons, and there will be minor variations in the geometry upon string melting in \textsc{AMPT}.   However,
reasonable variations in the smearing parameter $\sigma = 0.4$ fm cannot resolve the full differences.   
 


\begin{figure*}[ht]
\centering
\includegraphics[scale=0.88]{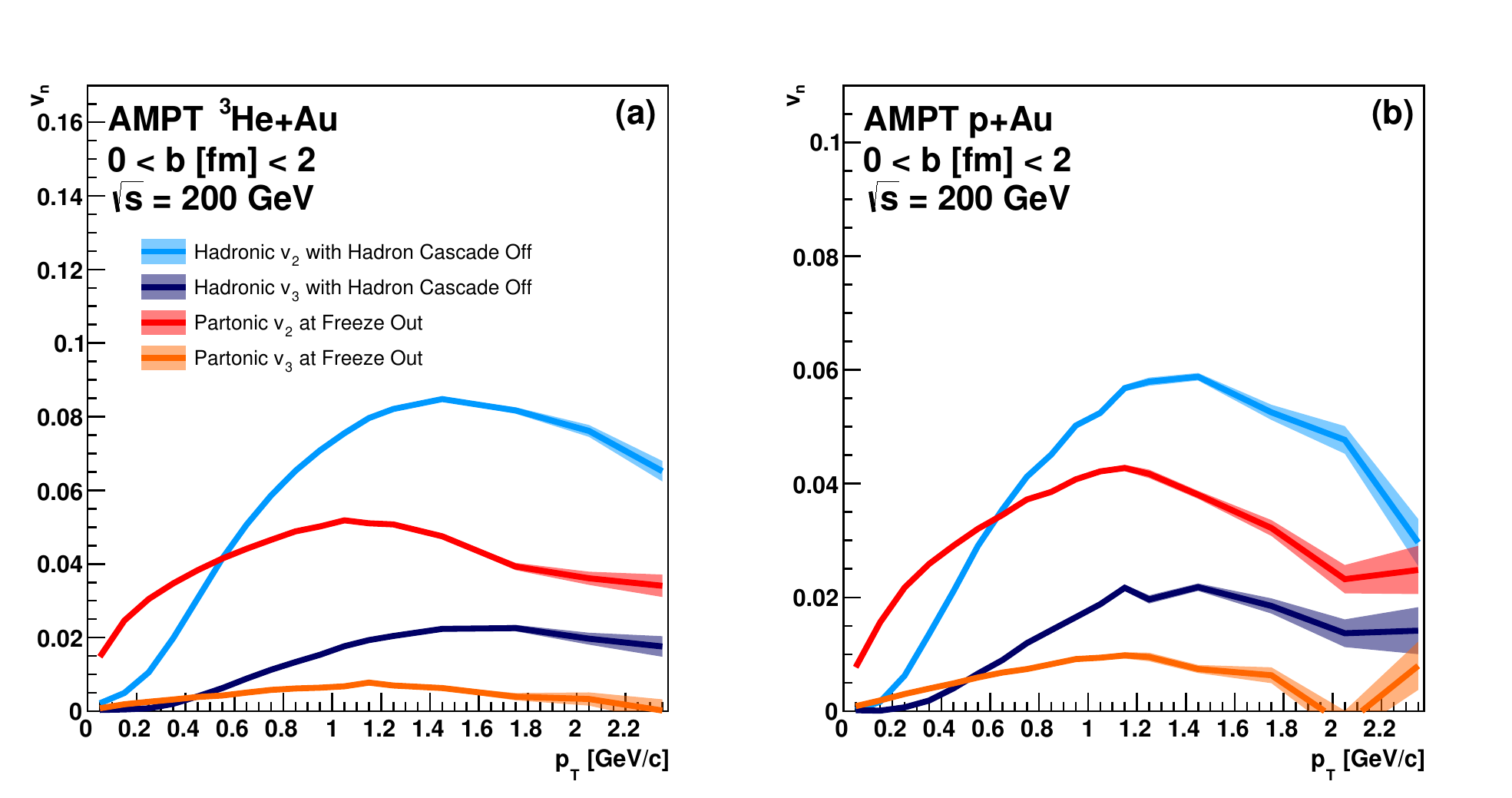}
\caption{Elliptic and triangular anisotropy moments in (a) \hau and (b) \pau collisions following partonic scattering using partons immediately prior to, and hadrons immediately after hadronization via coalescence.}
\label{fig_freezeout}
\end{figure*}

The $v_2$ in \pau is predicted to be substantially smaller than in \dau and \hau and the $v_3$ in \hau is predicted
to be substantially larger than in \pau and \dau, both from simple geometry.
Although we have qualitatively confirmed the expected scaling of momentum anisotropy with initial geometry across a variety of small collision systems, the exact mechanism responsible for the anisotropies and the deviations from geometry scaling,
for example at low \pt, must lie within the inner workings of the \textsc{AMPT} model. 

Several authors have identified partonic scattering as the critical mechanism for the development of $v_n$ in this context~\cite{He:2015hfa,bzdak_elliptic_2014,ma_long-range_2014}. However, partonic scattering is only the first stage in the evolution of the partons that emerge from string melting, and it is relevant to examine the effects that hadronization via coalescence and the subsequent hadron cascade have on the final $v_n$ values. 

To that end, we repeat our calculations for \hau and \pau, enabling and disabling the hadronic cascade, and effectively turning off partonic scattering by setting the parton cross-section to very nearly zero. The results are shown in Figure \ref{fig_cascade}. The first noteworthy feature in the top panels, corresponding to \hau, is that keeping the partonic scattering phase with $\sigma=1.5$ mb, but turning off the hadron cascade has a substantial effect on $v_n$. In fact, the blue and violet curves show that a late-stage hadron cascade actually increases $v_2$ by about 20\% for \pt$>1$ GeV/c and by roughly 100\% for \pt$<$ 0.5 GeV/c. This effect is even more pronounced for $v_3$. Additionally, the green curves show that even when the partonic phase is turned off, a sizable $v_2$ and $v_3$ still develop by  virtue of final-state hadronic interactions, with the effect being more pronounced for $v_3$. Finally, as a consistency check, disabling both the parton and the hadron cascades results in the collapse of $v_n$, as shown by the orange curves.

The same analysis is carried out for \pau in the bottom panels of Figure \ref{fig_cascade}. We observe that including a hadron cascade after the partonic scattering phase has a much smaller effect on the measured $v_2$ or $v_3$. Furthermore, the $v_n$ that develops from the hadron cascade alone when turning off partonic scattering is much less substantial than in the case of \hau, as evidenced by the green curves, showing that in \pau the bulk of the $v_n$ originates in the parton cascade.

We now examine the role of hadronization in the development of $v_n$. Figure \ref{fig_freezeout} shows $v_2$ and $v_3$ calculated for \hau and \pau collisions with a partonic scattering phase using $(i)$ partons at freeze out and $(ii)$ hadrons immediately after coalescence. For high \pt, we observe that hadronization increases $v_2$ and $v_3$ in both collision systems. However, for \pt$<$ 0.5 GeV/c, the effect of hadronization is to reduce $v_n$, as evidenced by the crossing of the curves in the figure. This can be understood in terms of quark coalescence dynamics. Since hadrons are produced by aggregating partons in spatial proximity and with collimated momenta, the coalescence yields hadrons with transverse momentum greater than that of their constituent quarks, hence increasing $v_n$ at higher \pt. It is notable that this effect is greater in \pau than in \hau.

The detailed mechanism and its relation to the number of parton scatterings in the early \textsc{AMPT} stage still require further
elucidation.   That said, it is clear that the \textsc{AMPT} coalescence prescription and following hadronic cascade substantially
modify and amplify this early stage effect. Therefore, the deviations from geometric scaling shown in Figure~\ref{fig_vn_ratio} 
appear to arise from the relative dominance of these stages as a function of \pt and collision system.

\section{Summary}
Recent intriguing experimental observations at RHIC and the LHC have raised the question of whether small droplets of QGP 
can be formed in small collision systems. From among several competing models, nearly inviscid hydrodynamic calculations, both at RHIC and the LHC, give reasonable account of the measured anisotropy coefficients. 
However, parton scattering in transport models---\textsc{AMPT} with string melting, in particular---has also been shown to provide an adequate description of the long-range azimuthal correlations and momentum anisotropy coefficients measured in \pp and \ppb at LHC energies. In this paper, we extend these calculations to RHIC energies, focusing on the insight that can be gained by varying the initial geometry of the projectile nucleus. 

We find that \textsc{AMPT} is capable of reasonably reproducing the measured elliptic and triangular flow coefficients for central \dau  and \hau collisions at \sqsn = 200 GeV for $\pt <1 \text{ GeV/c}$. However, \textsc{AMPT} underestimates the measured values for higher \pt. With this observation, we ascertain the validity of the model for rendering initial geometric anisotropy into final-state particle momentum correlations for small systems at both the RHIC and LHC energy scales. We also make predictions for elliptic and triangular anisotropy coefficients in \pau collisions at \sqsn = 200 GeV and qualitatively relate these results to calculated initial-state geometric anisotropy. 

However, we also find that partonic scattering is not the only source of the substantial elliptic and triangular momentum anisotropies in the \textsc{AMPT} model. Hadronization and the subsequent hadron cascade exert important modifications on the $v_n$ from partonic scattering, with strong dependences both on \pt and the intrinsic initial geometry of the system.
Direct comparisons with experimental data in these new systems, with both \textsc{AMPT} and various hydrodynamic models, is anticipated to shed light on the physical dynamics involved in these collision systems. To finalize, we highlight the need to identify additional observables that provide a more stringent discrimination between initial geometry and its translation to final-state correlations. 

\begin{acknowledgments}

We acknowledge funding from the Division of Nuclear
Physics of the U.S. Department of Energy under Grant
No. DE-FG02-00ER41152.   We also thank Paul Romatschke, Paul Stankus, and Shengli Huang for useful discussions.
\end{acknowledgments}



\bibliography{ampt_rhic}   

%


\end{document}